\documentclass[%
 aip,
rsi,
 amsmath,amssymb,
 reprint,%
]{revtex4-1}

\usepackage{graphicx}
\usepackage{dcolumn}
\usepackage{bm}

\usepackage[utf8]{inputenc}
\usepackage[T1]{fontenc}
\usepackage{mathptmx}
\usepackage{etoolbox}

\makeatletter
\def\@email#1#2{%
 \endgroup
 \patchcmd{\titleblock@produce}
  {\frontmatter@RRAPformat}
  {\frontmatter@RRAPformat{\produce@RRAP{*#1\href{mailto:#2}{#2}}}\frontmatter@RRAPformat}
  {}{}
}%
\makeatother
\begin{document}
\begin{widetext}
The following article has been accepted by \textit{Review of Scientific Instruments}. After it is published, it will be found at \verb|https://publishing.aip.org/resources/librarians/products/journals/|.

\newpage
\end{widetext}

\preprint{AIP/123-QED}

\title[]{ Optical stabilization for laser communication satellite systems through proportional–integral–derivative  (PID) control and reinforcement learning approach}
\author{A. Reutov}
 \email{ar@rootml.com}
 \altaffiliation[Also at ]{Moscow Center for Advanced Studies, Moscow, Russia}
\author{S. Vorobey}%
\author{A. Katanskiy}%
\author{V. Balakirev}%
\affiliation{LLC Science Trends, 119331 Moscow, Russia
}%

\author{R. Bakhshaliev}
\author{K. Barbyshev }
\author{V. Merzlinkin}
\author{V. Tekaev}

\affiliation{LLC QSpace Technologies, 121205 Moscow, Russia
}%

\date{\today}

\begin{abstract}
    One of the main issues of the satellite-to-ground optical communication, including free-space satellite quantum key distribution (QKD), is an achievement of the reasonable accuracy of positioning, navigation and optical stabilization. Proportional–integral–derivative (PID) controllers can handle with various control tasks in optical systems. Recent research shows the promising results in the area of composite control systems including classical control via PID controllers and reinforcement learning (RL) approach. In this work we apply RL agent to an experimental stand of the optical stabilization system of QKD terminal. We find via agent control history more precise PID parameters and also provide effective combined RL-PID dynamic control approach for the optical stabilization of satellite-to-ground communication system.
\end{abstract}

\maketitle


\section{Introduction}
Quantum key distribution (QKD) is one of the most commercially successful quantum technologies. Based on the laws of quantum mechanics, it allows to securely share cryptographic keys over insecure public channels.
Free-space QKD shows itself as the flexible and effective approach of quantum communications.

The control, navigation and target designation tasks are widely represented in various fields of science and technology. For instance, guidance systems are used for various communication technologies, in particular for the free-space optical communication, both for classical communication and for QKD. Classical control approaches (e.g., proportional–integral–derivative (PID) control strategies) have firmly occupied a place in many control tasks. Nevertheless, modern experimental methods of reinforcement learning \cite{Sutton1998} were applied in such complex and non-trivial areas as heating, ventilation and air-conditioning (HVAC) control \cite{Wei, Energy} or self-driving cars \cite{ Driving2017, Driving2021}. However, reinforcement learning is not a panacea and may not always significantly outperform classical control systems \cite{Bohn2019}. But is it possible to combine classical control methods and machine learning, what results will this lead to?

In this paper, we aim to improve existing equipment for free-space quantum communications. We explore the reinforcement learning (RL) algorithm for tuning of the PID controller parameters. The system can be controlled effectively in different modes using different optimal sets of PID parameters. Finding of this sets that are optimal within the boundaries of different system modes can be solved by RL approach. This leads to possible better result of PID-RL combination (PID controller retuned by RL agent due to change of system states) than simple PID controller with one set of parameters. Furthermore, instability may occur for the direct RL approach (especially in the early stages of learning). While a PID controller can provide basic stability for a wide range of coefficients, the RL agent can control PID parameters more safely.

The PID device are used for control of a micro-electromechanical system with mirror (MEMS mirror) in optical stabilization systems of laser beam. 
We work with stabilization system similar with ones applied to the receiver and the transmitter of the satellite-to-ground quantum key distribution \cite{Khmelev2021}.
Several articles \cite{DRPID,Self-preserving, MIMOPID} investigate the combination of PID control with RL and show better results than application of baseline PID controllers and than a RL agent that directly controls an object of influence. Combination of reinforcement learning algorithm and PID have been applied to simple toy models (e.g., holding a pendulum at unstable equilibrium point \cite{DRPID}) as well as to more realistic scenario \cite{Self-preserving}, where the control object was described by a second-order plus dead-time model. The main tool for these studies is RL algorithm suitable for continuous parameter control.

In our work, we follow \cite{DRPID,Self-preserving, MIMOPID} in RL model selection and also use Deep Deterministic Policy Gradients (DDPG) algorithm \cite{DDPG1}. It is common to use a software simulation as an environment for RL tasks. We also present an experimental stand that realistically reproducing an optical stabilization system of the satellite QKD receiver scheme presented in \cite{Khmelev2021}, in particular, the movement of a MEMS mirror. The control task is to keep the light spot at a given point of the sensor under the simulated external disturbance. There are random time delays between the agent and the control object due to the imperfections: we used the User Datagram Protocol (UDP) to communicate with the stand and this protocol suffers from such delays. Hence, we have a more complicated statement of the problem than common for RL problem \cite{Sutton1998}: there are unequal time intervals between the actions of the RL agent and they should be taken into account. It leads to the hypothesis that the control process in our task are represented as a semi-Markov process \cite{Sutton1999Semi-Markov}, not a Markov one. The last one is more typical for RL research \cite{Sutton1998}. During our investigation we find more optimal PID controller parameters through analysis of the training data of the RL agent and show their better performance compared to baseline PID controller settings which are tuned via Ziegler–Nichols tuning method \cite{Ziegler1942}. We also present the agent's dynamic control of the PID parameters by RL agent outperformed baseline PID.

Our article is organized as follows:
\begin{itemize}
     \item In Section \ref{sec:model} we formulate the reinforcement learning problem.
     \item Section \ref{sec:exp_part} is dedicated to the description of the experimental stand, the results of training of the RL agent and tests carried out on it.
     \item Finally, Section \ref{sec:discussion} concludes the paper with a discussion and a brief summary.
\end{itemize}

\section{Reinforcement learning model}\label{sec:model}

For the reinforcement learning problem, we consider a feedback system consisting of an agent and an environment. The environment is described by state $s_t$, the agent acts on the environment by action $a_t$, thereby changing state of the environment to $s_{t`}$ and receiving reward $r_t$. A process of transition from state $s_t$ to state $s_{t'}$ is modeled as a Markov (or semi-Markov) decision process. Policy $\pi(s) = a$ denotes a mapping of a state space $\{s_t \}$ into an action space $\{a_t\}$. Policy allows the agent to choose action $a_t$ for a given state $s_{t}$ and represents agent`s behaviour. An optimal policy  maximizes sum of rewards $\sum_{t} \gamma^{t} r_{t}$. The Bellman equation enables to find the optimal policy and provide theoretically convergence to it. One of the simplification of this equation is given by:
\begin{equation}
    Q(a_t, s_t) = r_t + \gamma \underset{a_{t`}}{\rm max}[Q(a_{t`},s_{t`})],
    \label{eq:Bellman}
\end{equation}
where the quantity $\gamma$ is the discount factor ($\gamma \in (0,1)$) and $Q(a_t, s_t)$ is the Q-value defining the policy $\pi(s)$: with given state $s_t$, the Q-value $Q(a_t, s_t)$ specifies the weights for the possible actions $a_t$. For example, a policy $\pi(s)$ can be constructed by choosing an action that maximizes $Q(a_t, s_t)$ in a given $s_t$.

Our task is development of adaptive RL tuning of PID controllers, i.e. PID controllers are constantly retuned by the RL agent (depending on new observable data).
We formulate the problem of adaptive tuning in our case of PID control over the mirrors of the optical stabilization system in terms of the reinforcement learning problem as follows:
\begin{itemize}
     \item \textbf{State} is a set of several $k$ \textit{observations}: $s_t = \{o_{t,i}\}_{i=0}^k$. Each observation $o_{t,i}$ is a set of the value of the current position ($x_{{\rm pos},i}$ and $y_{{\rm pos},i}$) of the bright spot on the receiving sensor, the output from the PID controllers ($u_{x,i}$ and $u_{y,i} $) and target position ($x_{\rm target}$ and $y_{\rm target}$). A target position is changed with every new state.
     \item \textbf{Action} $\pi(s_t) = a_t$ is a set of the new parameters of the PID $\{P_x,\,I_x\}$ and the PID $\{P_y,\,I_y\}$ controlling the mirror along the axes $X$ and $Y$ axis respectively. We do not change the differential parameters $D_{x,y}$, leaving they equal $D_{x,y}^{\rm baseline}$ from the baseline coefficient set (see Table \ref{table:baseline_test}).
     \item \textbf{Reward} $r_t$ is sum of distances between the positions $(x_{{\rm pos},i},\,y_{{\rm pos},i})$ and the target position $(x_{\rm target},\,y_{\rm target})$:
     \begin{equation}
         r = c_r \sum_i \sqrt{(x_{{\rm pos},i} - y_{{\rm pos},i})^2+(y_{\rm pos} - y_{\rm target})^2},
     \end{equation}
     where $c_r$ is the normalization factor. We additionally penalize the agent by value $r_{\rm out \, baseline}$ if its reward $r$ significantly exceeded the reward obtained by choosing the baseline action $a_{\rm baseline}$ which is a known set of PID parameters $\{P_x^{\rm baseline}, \,I_x^{\rm baseline},\,D_x^{\rm baseline}\}$ and $\{P_y^{\rm baseline},\,I_y^{\rm baseline},\,D_y^{\ rm baseline}\}$ (Table \ref{table:baseline_test}).
     \item We also take into account the time difference $\Delta_t$. This difference is proportional to the time intervals between actions performed, and the Bellman equation is rewritten in the same manner as presented in \cite{Sutton1999Semi-Markov}:
     \begin{equation}
     Q(a_t, s_t) = r_t + \gamma^{\Delta_t} \underset{a_{t`}}{\rm max}[Q(a_{t`},s_{t`})],
     \label{eq:Bellman_new}
\end{equation}
\end{itemize}

To vary the agent's experience and explore the action and state spaces the $\epsilon$-greedy policy is used, which adds a randomness to the actions $\pi(s)$:
\begin{equation}
    a_{\rm actual} =\pi(s)*x_{\rm OU}(\epsilon)
    ,
\end{equation}
where $x_{\rm OU}(\epsilon)$ is a stochastic Ornstein–Uhlenbeck noise \cite{Uhlenbeck1930}, which was commonly added in similar control tasks (e.g. \cite{MIMOPID}):
\begin{equation}
    x_{\rm new} = x_{\rm old} + \Delta t \frac{\mu- x_{\rm old}}{\tau} + \epsilon \sqrt{\frac{2 \Delta t}{\tau}} \cdot U(0,\,1) \,.
\end{equation}
Here $U(0,\, 1)$ is the uniform continuous distribution in the interval $(0,\, 1)$, parameters of this process set as $\mu=1$, $\tau = 0.15$ and $\theta = \epsilon$, and exploration parameter $\epsilon$ is discounted from $\epsilon_i=0.2$ to $\epsilon_f = 0.0001$.

We use the Deep Deterministic Policy Gradient (DDPG) model \cite{DDPG2} as a learning agent following previous research in this field \cite{DRPID, Self-preserving, MIMOPID}. This choice is made due to its simplicity and applicability to the class of continuous problems \cite{NIAN2020, Henderson2018}. Deterministic policy was also chosen due to the need to have stable and predictable source of actions. The DDPG develops from a combination Deep Q-learning Network approach (more common for discrete problems) and policy gradient method. DDPG has an Q-value estimator determined by a neural network $Q(s,\, a)$ which is called ``critic''. This part of algorithm takes the input pair $(s_t,\, a_t)$ and outputs Q-value. The policy is set by another neural network $\pi(s)$ called ``actor'' for which the critic is used as the loss function for weight update (detailed description of algorithm presented in an original article\cite{DDPG1}). 

We trained DDPG due to an interaction with an experimental stand (see description of the stand in Section \ref{sec:stand}), without pretraining on program simulation and with randomly initialized weights of the critic and actor neural networks. We define neural networks inside DDPG model as multilayer perceptrons with two hidden layers and choose their size as $100$ and $60$ neurons for first and second hidden layer respectively.
Our choice of activation layers, hyperparameters and an optimizer coincided with those presented in the original DDPG article \cite{DDPG1}. This set of hyperparameters, chosen activation layers and an optimizer have benchmark testing \cite{Henderson2018} and were previously applied for PID control tasks \cite{Self-preserving,MIMOPID}.

\section{Experiment and results}\label{sec:exp_part} 

\subsection{Experimental setup}\label{sec:stand}

\begin{figure}[h!]
\centering
\includegraphics[scale=0.5]{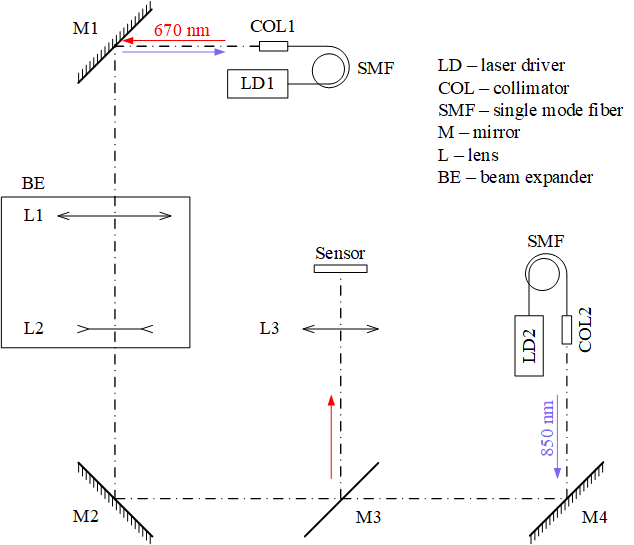}
\caption{Optical scheme of the stand for the laser beam optical stabilization system.}
\label{fig:Stand}
\end{figure}

The optical scheme is shown in Fig. \ref{fig:Stand}. The light of the laser diode LDI-670-FP-20-B (LD1) with a wavelength of 670 nm is sent to a movable fast piezoelectric tip-tilt platform S-330 Physik Instrumente with mirror (M1) using an optical collimator F280FC-B (COL1), which provides parallel beam paths. The movable mirror (M1) simulates external disturbance. It leads to imperfect coupling of laser light into the transceiver optical system. The light reflects from the M1 mirror and passes through the Galileo GBE03-B (BE) afocal telescopic optical system, which reduces the beam diameter. Then, the light reflects from the Mirrorcle Technologies MEMS mirror (M2) which is needed to compensate external disturbance and for coupling laser light into the center of the sensor sensor of the precise guidance system. After the MEMS mirror there is a dichroic optical element DMLP805 (M3) on the beam optical axis, which reflects light towards a lens that focuses the laser beam on the sensor VC MIPI IMX392 (Sensor).

Mirrors M1 and M2 are controlled by a computer via a Raspberry PI connected to the sensor sensor. The digital PID controller on the Raspberry PI generates voltage values that are applied to the MEMS mirror and are proportional to the rotation angles. A remote computer connected to the Raspberry PI via UDP generates voltage values for the piezoelectric mirror (M1): this voltage values simulate external disturbance and are proportional to displacement angle of M1. The computer gets data from the Raspberry PI, received from the sensor and digital PID controller, and sends parameters for the controller and new target values to adjust the position of the laser spot on the sensor. The sensor has the measurement frequency about $120$ Hz and our RL control system operates and trains at this frequency in real time both.

\begin{figure}[h!]
\centering
\includegraphics[scale=0.37]{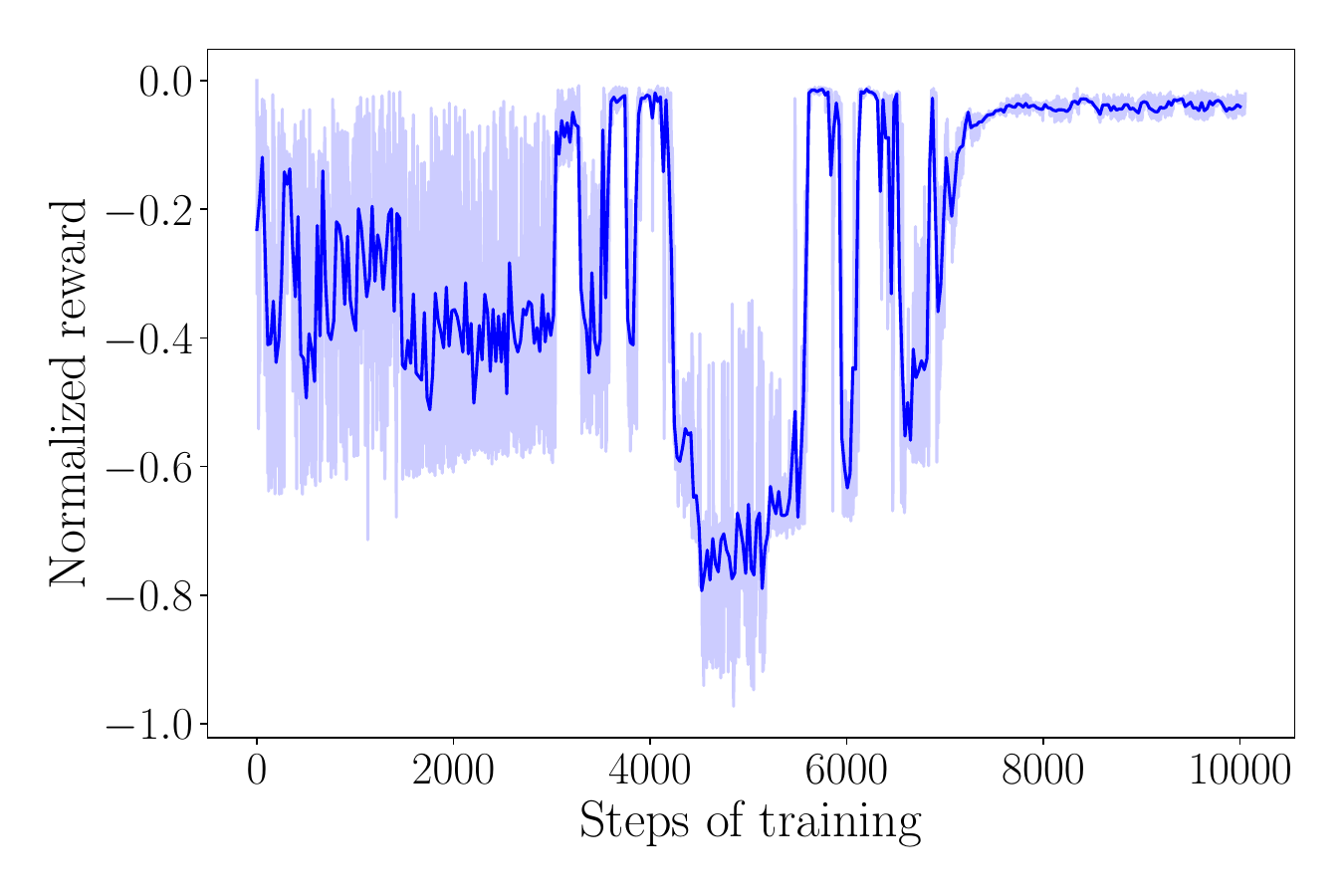}
\caption{Reward $r_t$ for $10000$ learning steps (bright blue line). The dark blue line shows the reward averaged over the last $10$ steps.}
\label{fig:reward}
\end{figure}

\subsection{Training on the stand and tests}
The training process for our DDPG implementation continues $10000$ steps (see Fig. \ref{fig:reward}). During training we change the values of $x_{\rm target}$ and $y_{\rm target}$ to a new value shifted by $100$ or $-100$ pixels randomly:
\begin{equation}
\begin{split}
x_{\rm target} \rightarrow{} x_{\rm target} \pm 100\\
y_{\rm target} \rightarrow y_{\rm target} \pm 100  
\label{eq:train_dist}
\end{split}
\end{equation}
Using a step function when adjusting the PID controller is a universal method for analyzing the system's response to short-term pulse disturbances in a wide frequency range.
\begin{table}[ht]
\begin{center}
  \begin{tabular}{|c|c|c|c|c|c|c|}
  \hline
  &$P_x$&$I_x$&$D_x$&$P_y$&$I_y$&$D_y$\\
  \hline
  $a_{\rm baseline}$&10 &5 &0 &16 &6.8 &2.5\\
  \hline
  $a_{\rm test}$&15& 15.3 & 0 & 13& 13.8 & 2.5 \\
  \hline
  \end{tabular} 
\end{center}
  \caption{Baseline and test sets of PID parameters.}
  \label{table:baseline_test}
\end{table}

We assumed that RL-PID approach provide different sets of PID parameters in different possible system modes (previously, manual tuning showed several possible set optimal in different domains of the system state). However, the agent's actions converges into the small domain of possible action values.
We choose the interval of training from $k = 8000$ to $N=10000$ steps as the interval where the agent achieves successful training and system demonstrates stable reward feedback. Based on the data of the actions executed by agents in the interval $(8000, \,10000)$, we define a set of actions $a_{\rm test}$ (see Table \ref{table:baseline_test}). These actions are the averaged sum of actions $\sum_{i=k}^N a_i /(N-k)$: here we assume that the agent's actions $a_k$ can converge to a constant optimal value $a_{\rm opt}$ and we choose $a_{\rm test} \approx a_{\rm opt}$. 

The tests were organized as follows. We replace the simulation of external disturbance with new one: the angles of the mirror (M1) of the experimental stand are changed according to the Wiener process $W_{x,(y)}(t)$ (simulates external disturbance):
\begin{equation}
    (x_{\rm target},\, y_{\rm target}) = (W_x(t), W_y(t))
    \label{eq:test_dist}
\end{equation}
Such simulating an external disturbance in accordance with the Wiener process is the closest to the real operating conditions of the stabilization system. It is worth noting that the agent was previously trained on stepwise change of target values (\ref{eq:train_dist}) and the agent did not have access to experience with such behavior of target values. It allows to check the stability of DDPG when conditions of environment and control task change. 

The tested control action policy are applied to the stand. During the test, we collected statistics of the light spot position on the sensor sensor over $1000$ steps. In total, we tested: a baseline set of parameters $a_{\rm baseline}$ (Table \ref{table:baseline_test}); set of parameters $a_{\rm test}$ (Table \ref{table:baseline_test}); sets obtained from the output of the trained actor neural network $\pi(s)$; as well as sets obtained from the output of the actor neural network $\pi(s)_{\rm retrain}$ from the training model additionally retrained. In the last case, we use a trained DDPG at the start of testing and additionally run the process of training and update the weights of the DDPG networks during the test. For our analysis, we consider the quantities $\Delta$, $\Delta_x$ and $\Delta_y$, which are given as:
\begin{equation}
    \begin{split}
        \Delta &= \sqrt{(x_{\rm pos}-x_{\rm target})^2+(y_{\rm pos}-y_{\rm target})^2},\\
        \Delta_x &= |x_{\rm pos}-x_{\rm target}|,\\
        \Delta_y &= |y_{\rm pos}-y_{\rm target}|.
        \nonumber
    \end{split}
\end{equation}
This values represent the distance between target and actual position of light spot and the absolute coordinate differences.
Mean values $\overline \Delta$, $\overline \Delta_x$, $\overline \Delta_y$ and standard deviations $\sigma_\Delta$, $\sigma_{\Delta_x}$, $\sigma_{\Delta_y}$ are presented in Table \ref{tab:all_results} for all 4 cases (baseline set $a_{\rm baseline}$, test set $a_{\rm test}$ and tested networks $\pi(s)$ and $\pi(s)_{\rm retrain}$). 

All test actions gives better result (smaller distance between actual position and target position) than known baseline PID parameters $a_{\rm baseline}$. According to Table \ref{tab:all_results}, the best score (minimal deviation of beam position) are achieved with retrained during testing network  $\pi(s)_{\rm retrain}$. We assume, that new conditions (new target position pattern) can be tough for the agent and including new experience into the agent behaviour via training helps two obtain the best result. 
\begin{table}[ht]
\begin{center}
  \begin{tabular}{|c|c|c|c|c|c|c|}
  \hline
  &$\overline \Delta$& $ \sigma_\Delta$ & $\overline \Delta_x$ & $\sigma_{\Delta_x}$ & $\overline \Delta_y$ & $ \sigma_{\Delta_y}$\\
  \hline
  $a_{\rm baseline}$&7.83 & 5.48 & 4.96 & 5.16& 4.97& 3.9\\
  \hline
  $a_{\rm test}$&6.42&3.78&3.48&2.74&4.63&3.79\\
  \hline
  $\pi(s)$&7.18&4.51&3.63&2.8&5.35&4.72\\
  \hline
  $\pi(s)_{\rm retrain}$&\textbf{6.12}&4.69&\textbf{3.12}&2.58&\textbf{4.53}&4.75\\
  \hline
  \end{tabular}
\end{center}
  \caption{
  Means $\overline \Delta$, $\overline \Delta_x$, $\overline \Delta_y$ and standard deviations $\sigma_\Delta$, $\sigma_{\Delta_x}$, $\sigma_{\Delta_y}$. All values are given in pixels of the receiving sensor ($1 {\rm px} = 3.45 {\mu m}$). The smallest means $\overline \Delta$, $\overline \Delta_x$ and $\overline \Delta_y$ are achieved when there is used $\pi(s)_{\rm retrain}$ from the model retrained during the test.}
  \label{tab:all_results}
\end{table}

\section{Summary}\label{sec:discussion} 

This work proposes the application of reinforcement learning for a laser beam optical stabilization in satellite optical communication systems. The presented method of control and tuning using RL are assumed to apply for an increasing of secrete key generation in satellite QKD systems. We use the DDPG algorithm both for PID controller tuning and for adaptive PID control. The best test results are achieved for the adaptive approach via $\pi(s)_{\rm retrain}$: the average deviation $\overline \Delta$ of the light spot on the receiving sensor is reduced by $21.7 \pm 3.2 \%$ compared to that are obtained by PID control with the manually set parameters $a_{\rm baseline}$ (reduction from $7.83$ pixels to $6.12$ pixels on the average). The new $a_{\rm test}$ parameters set using DDPG shows a $17.9 \pm 3.2 \%$ reduction compared to $a_{\rm baseline}$ (reduction from $7.83$ pixels to $6.42$ pixels on the average).

Our future research is aimed at extending the proposed approach from the experimental stand to the actually working satellite optical communication systems. We plan to use adaptive optics at our ground station, including with RL control, for wavefront correction \cite{zou:hal-04605102} and, as a result, increase the efficiency of the communication link. In addition, some types of communication systems such as satellite quantum key distribution systems  not directly require a maximum accuracy of laser beam placement: the minimization of QBER and the maximization of secret key rate are objectives in the QKD practical systems potentially optimized by RL methods for optical stabilization.

\hfill

\begin{acknowledgments}
The authors would like to acknowledge Liubov Mashkovtseva for thoughtful and productive
discussions.
\end{acknowledgments}

\section*{Data Availability Statement}
The data that support the findings of this study are available from the corresponding author upon reasonable request.

%

\end{document}